# THE LOGARITHMIC CURVATURE GRAPHS OF GENERALIZED CORNU SPIRALS

[1]R.U. Gobithaasan, [2]J.M. Ali & [3]Kenjiro T. Miura
[1] Dept. of Mathematics, FST, University Malaysia Terengganu, 21030 Kuala Terengganu, Malaysia
[2] School of Mathematical Sciences, University Sains Malaysia, 11800 Minden, Penang, Malaysia.
[2] Graduate School of Science and Technology, Shizuoka University, Japan.

e-mail: gr@umt.edu.my & jamaluma@cs.usm.my, tmkmiur@ipc.shizuoka.ac.jp

**Abstract.** *The Generalized Cornu Spiral (GCS) was first proposed by Ali et al. in 1995 [9]. Due to the monotonocity of its curvature function, the surface generated with GCS segments has been considered as a high quality surface and it has potential applications in surface design [2]. In this paper, the analysis of GCS segment is carried out by determining its aesthetic value using the log curvature Graph (LCG) as proposed by Kanaya et al.[10]. The analysis of LCG supports the claim that GCS is indeed a generalized aesthetic curve.*

## 1. Introduction

Geometric modeling deals with the study of free-form curve as well as surface design and it is one of the most basic foundations in the product design environment. The process of information extraction from a geometrical model is known as shape interrogation technique [1], such as the inspection of a curvature profile. Shape interrogations are vital in manufacturing in order to verify whether the designed product meets its functionalities and aesthetic shapes. Since the foundation of curve development for CAD systems revolves around the flexibility of free form curve construction, its curvature which indicates the shape aesthetics has been overlooked. Unlike spirals, controlling the curvature of the stated curves has been a major problem. Curvature controlled curves are not only vital for product design but also play an important role particularly in robot trajectories, railway and highway design.

In 1972, Nutbourne [3] developed a technique known as curve synthesis, where planar curves are generated by integrating the curvature profile functions. There have been many interests in carrying out curve synthesis for linear curvature functions whereby segments of Cornu spiral or clothoid are generated. Pal & Nutbourne [4], Schechter [5] and Meek & Walton [6] have carried out detailed investigation on curvature function consisting of piecewise linear functions.

In 1995, Ali [9] proposed an alternative curvature function in the form of a rational linear function for curve synthesis. The resultant curve is a new family of spiral which has been denoted as generalized cornu spiral or GCS. Cripps [12] stated that GCS is more versatile than clothoid in terms of shape description. It is also considered as a quality curve since its curvature function is monotonic [2, 3].

Harada et al. [11] proposed a different kind of approach to analyze the characteristics of planar curves used for automobile design. The relationship between the length frequencies of a segmented curve with regards to its radius of curvature is plotted in a log-log coordinate system and is called Logarithmic Distribution Diagram of Curvature (LDDC). Two circular arcs with the same radius but different length would generate similar curvature profile; nevertheless LDDC would generate different shapes. To note, the generation of a LDDC is through quantitative methods where the user runs algorithmic steps to produce a histogram which illustrates the locus of the interval of radius of curvature and its corresponding length frequency. Furthermore, the needs the curve analysis experience in order to classify an arbitrary segment into appropriate number of classes to obtain the correct histogram which is a painstaking process.

In 2003, Kanaya et. al [10] proposed the generation of Logarithmic curvature Graph (LCG) to substitute LDDC. LCG is produces the exact/analytic relationship between the interval of radius of curvature and its corresponding length frequency. Hence, it suits well for measuring the aesthetic value of planar curves [7]. In 2006, Miura [8] proposed a generic formula to obtain planar curves with aesthetic properties named as Log Aesthetic Curves (LAC). Under certain conditions, LAC may represent natural spirals namely logarithmic spiral, cornu spiral and circle involute. Recently, Muira et. al reformulated LAC by means of variational principle and further applied this curve as a digital filter [15,16].

In this paper, we analyse GCS in terms of LCG in order to identify the aesthetic value of GCS. The measurement of aesthetic value of GCS has offered a positive insight for defining the properties that make a curve aesthetic. In the final section, numerical examples are shown for better comprehension of path that leads to defining high quality curves.

## 2. Curve Synthesis leading to GCS

### 2.1 General formulation for curve synthesis

Let $\boldsymbol{r}(s) = \{x(s), y(s)\}$, for $0 \leq s \leq S$ represents a planar curve with arc length parameterization and S corresponds to the total arc length of **r**(s). Unit tangent vectors **t**(s), unit normal vectors **n**(s) and the signed curvature κ(s) can be calculated directly as stated below:

$$\boldsymbol{t}(s) = \left\{\frac{dx(s)}{ds}, \frac{dy(s)}{ds}\right\} \; \boldsymbol{n}(s) = \left\{-\frac{dx(s)}{ds}, \frac{dy(s)}{ds}\right\} \text{ and } \kappa(s) = \left(\frac{dx(s)}{ds}\frac{d^2y(s)}{ds^2} - \frac{dy(s)}{ds}\frac{d^2x(s)}{ds^2}\right)$$

The curvature κ(s) is positive when the rotation is counter clockwise and vice versa. The signed angle θ(s) is defined in the interval of $0 \leq s \leq S$ and is measured in radians from the positive x axis to **t**(s): $\boldsymbol{t}(s) = \{\cos(\theta), \sin(\theta)\}$, where

$$\theta(s) = \theta(0) + \int_0^S \kappa(t)dt \tag{1}$$

The equation of the curve from its given curvature profile is [1]:

$$r(s) = \left\{x(0) + \int_0^s \cos\left[\theta(0) + \int_0^t \kappa(u)du\right] dt, \quad y(0) + \int_0^s \sin\left[\theta(0) + \int_0^t \kappa(u)du\right] dt\right\} \tag{2}$$

The integrals in equation (2) are the Fresnel integrals [17] which are difficult to evaluate directly, thus symbolic/numerical software can be utilized to generate points on the curve. Mathematica® has been used for the generation of figures and symbolic/numerical computing.

### 2.2 Examples of curve synthesis

A constant curvature function that produces a circular arc is the simplest form of curvature function for curve synthesis. A linear curvature function given by

$$\kappa(s) = a + bs, \; 0 \leq s \leq S \tag{3}$$

where a and b are constants corresponds to a segment of a clothoid and if equation (3) satisfies the given value of end curvatures, denoted by $\kappa_0$ and $\kappa_1$, then the corresponding curvature function is:

$$\kappa(s) = \left(1 - \frac{s}{S}\right)\kappa_0 + \frac{s}{S}\kappa_1 \tag{4}$$

Figure (1) shows an example of a segment of curve generated with linear curvature function where $\kappa_0 = 0$, $\kappa_1 = 2$ and S=1.

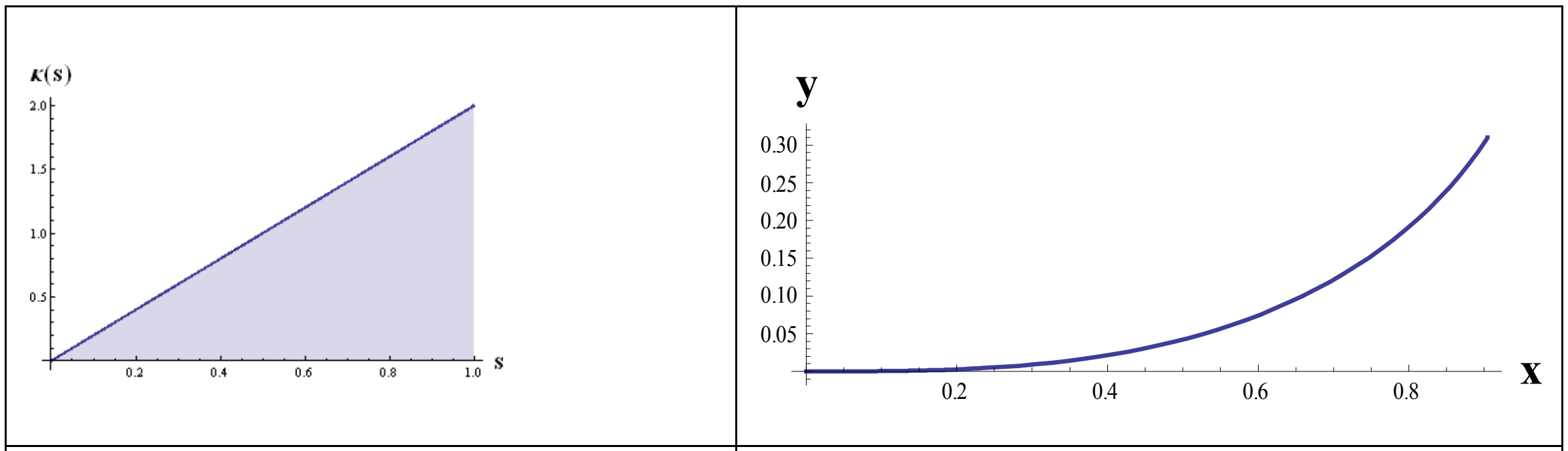

Linear curvature profile with where $\kappa_0 = 0$, $\kappa_1 = 2$ and S=1. | Corresponding curve segment (a segment of clothoid)

Figure 1. Curve Synthesis using linear curvature profile

Similarly, a quadratic curvature function given by $\kappa(s) = as^2 + bs + c$ (where a, b and c are constants) which satisfies the end curvature values can be written as:

$$\kappa(s) = as^2 + \frac{\kappa_1 - \kappa_0 - a\boldsymbol{S}^2}{\boldsymbol{S}} s + \kappa_0, \ 0 \le s \le \boldsymbol{S} \tag{5}$$

Since oscillations may occur in quadratic curvature function, it is not preferable. Equation (5) can further be modified with the introduction of total turning angle which can be derived from end tangents and given total arc length and one may obtain a somewhat acceptable curve with less oscillations. When a=0, a clothoid is produced but a quadratic curvature function cannot produce Logarithmic spirals which are considered aesthetic by nature and essential for aesthetic design intent [9].

### 2.3 Generalized Cornu Spiral (GCS)

Let a curve being defined in the interval $0 \le s \le S$, and its curvature function is represented by a rational linear curvature function as:

$$\kappa(s) = \frac{p+qs}{S+rs} \tag{6}$$

where $p, q, r$ and non negative $S$ are free variables of the curve segment. The resultant curve given by equation (2) is a family of Generalized Cornu Spiral denoted as GCS. It is noted that GCS will reduce to straight lines when $p = q = 0$, circular arcs when $r = q = 0$ , Logarithmic spirals when $q = 0$ and segments of clothoid when $r = 0$. To ensure that $\kappa(s)$ is well behaved and continuous over the stated interval the values of $r$ is restricted to $r > -1$. The inflection occurs at most once at $s = \frac{-p}{q}$.

Suppose the arc length of the GCS segment be S and if the end curvatures are $\kappa_0$ and $\kappa_1$, then we have $\kappa(0) = \kappa_0$ and $\kappa(S) = \kappa_1$. Thus, equation (6) produces the following set of equations [9]:

$$q = S\kappa_0 \tag{7}$$

$$pS + q = S(1+r)\kappa_1 \tag{8}$$

Solve p in terms of r, we get $p = (1+r)\kappa_1 - \kappa_0$. Finally, substituting p and q in equation (6), the curvature function becomes [9]:

$$\kappa(s) = \frac{(\kappa_1 - \kappa_0 + r\kappa_1)s + \kappa_0 S}{rs+S}, \ 0 \le s \le S \text{ and } r > -1 \tag{9}$$

By substituting equation (9) in equation (2), a GCS segment with specified arc length and end curvatures can be obtained. Depending on the selection for end curvatures, a GCS segment may have only one inflection point and the monotonicity of the curvature function is always preserved. These properties directly indicate that GCS segments are categorized as aesthetic and considered as a fair curve. In [12], Cripps created high quality surface patches by using GCS as the auxiliary curve. Through analyzing the aesthetic values of GCS, further insights for an alternative definition of aesthetic curves are proposed as compared to the definition proposed by Japanese researchers [7, 8 and 11].

## 3. Measuring the aesthetic values of GCS using LCG

In 2003, Kanaya et. al [10] proposed the generation of Logarithmic curvature Graph, abbreviated LCG and it is an analytical way of obtaining the relationship between the interval of radius of curvature and its corresponding length frequency. Thus, two curves with different length would generate distinct LCG regardless of the similarities of the shape of curvature profile. Gobithaasan et. al [7] proposed a constructive formula to obtain LCG and its gradient directly from a parametric equation. The analysis of the gradient of LCG is introduced as a tool which measures the aesthetic value of planar curves.

Let a planar curve defined as $c(t) = \{x(t), y(t)\}$ and its radius of curvature and arc length function is defined as $\rho(t)$ and $s(t)$ respectively. The LCG and its gradient for $c(t)$ can be obtained respectively [7]:

$$LCG(t) = \left\{ \log[\rho(t), \log\left[\frac{\rho(t)s'(t)}{\rho'(t)}\right]]\right\} \tag{10}$$

$$gradient(t) = 1 + \frac{\rho(t)}{\rho'(t)^2}\left(\frac{\rho'(t)s''(t)}{s'(t)} - \rho''(t)\right) \tag{11}$$

Note that the LCG is the graph of $\log\left[\frac{\rho(t)s'(t)}{\rho'(t)}\right]$ against $\log[\rho(t)]$ and the function $gradient(t)$ is the slope of this graph. The specification of radius of curvature($\rho_{GCS}(t)$) and arc length ($s_{GCS}(t)$) for GCS in order to obtain the LCG function and its gradient are as follows:

$$\rho_{GCS}(t) = \frac{rt+S}{(\kappa_1 - \kappa_0 + r\kappa_1)t + \kappa_0 S} \text{ and } s_{GCS}(t) = t \tag{12}$$

Hence, by substituting equation (12) and their derivatives into equation (10) and (11), the LCG graph and its gradient are obtained respectively:

$$LCG_{GCS}(t) = \left\{ Log\left[\left|\frac{rt+S}{(\kappa_1 - \kappa_0 + r\kappa_1)t + \kappa_0 S}\right|\right], Log\left[\left|\frac{(S + r\,t)\,(S\,\kappa_0 + t\,(-\kappa_0 + \kappa_1 + r\kappa_1))}{(1+r)\,S\,(\kappa_0 - \kappa_1)}\right|\right]\right\} \tag{13}$$

$$gradient_{GCS}(t) = \frac{((-1+r)\,S - 2\,r\,t)\,\kappa_0 + (1+r)\,(S + 2\,r\,t)\,\kappa_1}{(1+r)\,S\,(\kappa_0 - \kappa_1)} \tag{14}$$

Equation (14) can further be simplified and it is a linear function of t given as:

$$gradient_{GCS} = \left(\frac{2\,r\,(-\kappa_0 + \kappa_1 + r\,\kappa_1)}{(1+r)\,S\,(\kappa_0 - \kappa_1)}\right)t + \left(\frac{2\,r\,\kappa_0}{(1+r)\,(\kappa_0 - \kappa_1)} - 1\right) \tag{15}$$

Hence,

$m = \left(\frac{2\,r\,(-\kappa_0 + \kappa_1 + r\,\kappa_1)}{(1+r)\,S\,(\kappa_0 - \kappa_1)}\right)$ is the slope and $\left(\frac{2\,r\,\kappa_0}{(1+r)\,(\kappa_0 - \kappa_1)} - 1\right)$ is the vertical intercept.

Equation (15) indicates that the gradient of LCG is always a straight line. This element prevails as a constructive criterion for identifying aesthetic curves.

A curve is said to be an aesthetic curve if the gradient of LCG of the curve is constant or the LCG is represented as a straight line [7, 8 and 11]. However, by analyzing the LCG of GCSs, the findings indicate that the linearity LCG to represent aesthetic curves maybe too rigid. Since Japanese researchers [7, 8 and 11] indicated that aesthetic curves are high quality curves and Cripps [12] on the other end indicated that GCS is a high quality curve, the findings in this paper shows the possibility that the definition of a general classification of aesthetic curves is based on the following equation of LCG gradient:

$$gradient(s) = As + B \tag{16}$$

where A and B are arbitrary constants with s as its arc length. Based on Equation (16), the gradient becomes constant when A becomes zero, which is in accord with the findings of Japanese researchers. Equation (16) represents the gradient of GCS when $A \neq 0$. Hence, we believe that Equation (16) captures the essence of a general representation of high quality curves.

## 4. Numerical Examples

Figure (2) shows the curvature function of GCS with the values of r ranging from {100, 5, 2, 1, 0, -0.5,-0.9,-0.99} with $\kappa_0 = 0$, $\kappa_1 = 2$ and $S=\pi$.

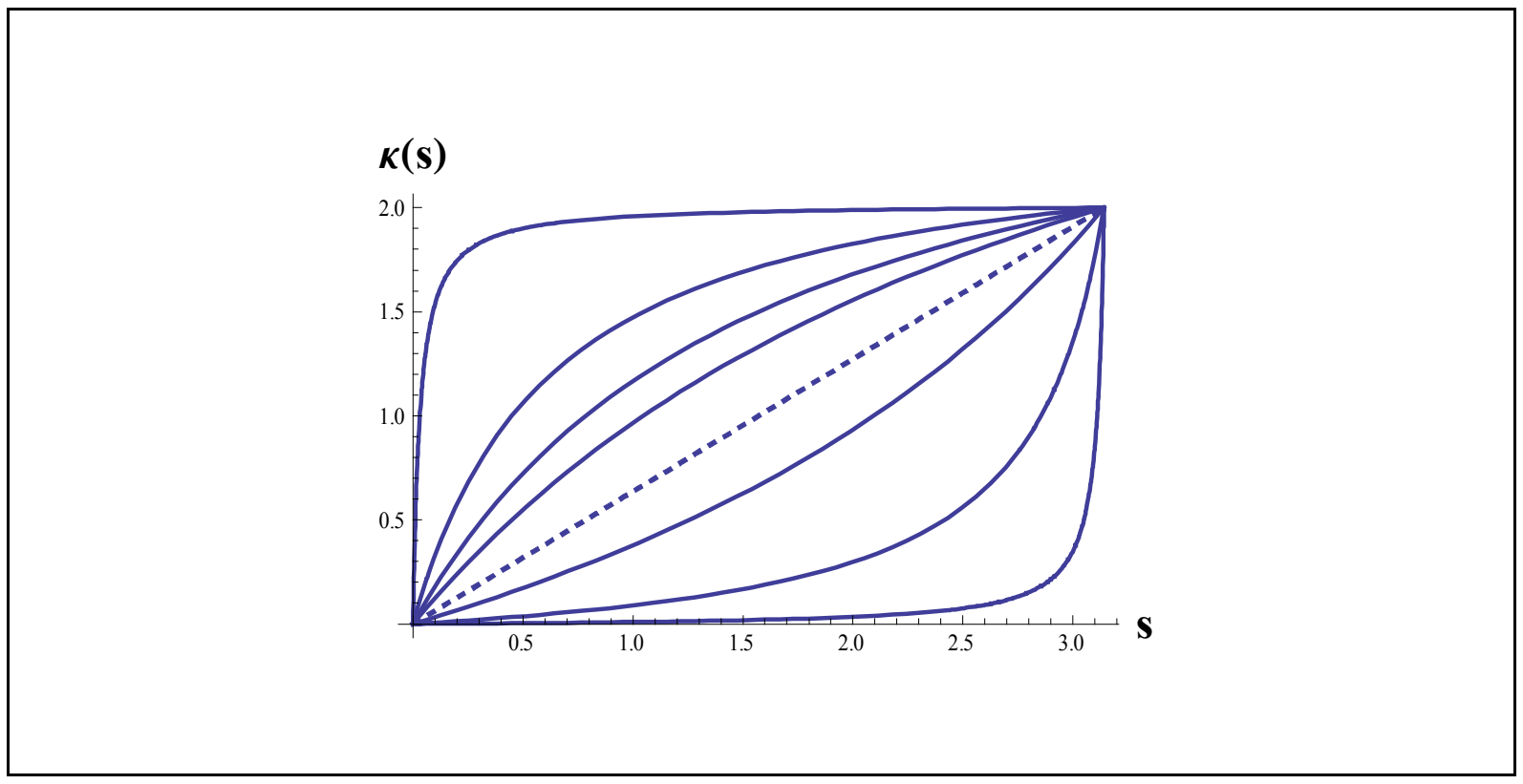

Figure 2. The dashed line is obtained when r=0 and the curvature function goes higher as r increases (r=1, 2, 5,100) and goes lower as r decreases (r=-0.5, -0.9, -0.99).

Figure (3) shows the corresponding GCS segments for the curvature profiles shown in Figure (2).

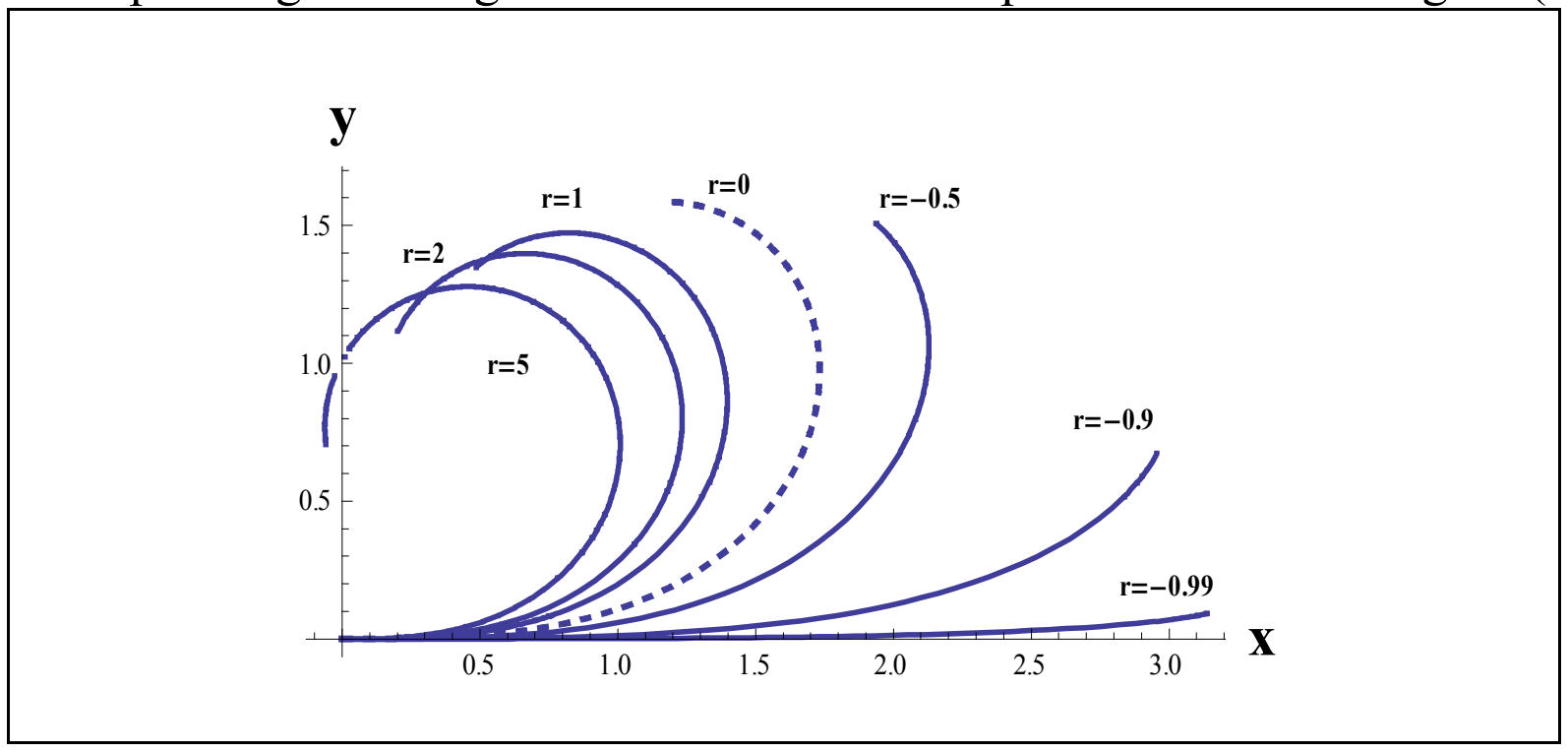

Figure 3. The dashed line is obtained when r=0 and the GCS segment curls to the left as r increases (r=1, 5) and flattens as r decreases (r=-0.5, -0.9,-0.99).

Figure (4) shows the corresponding LCG for GCS segments shown in Figure (3).

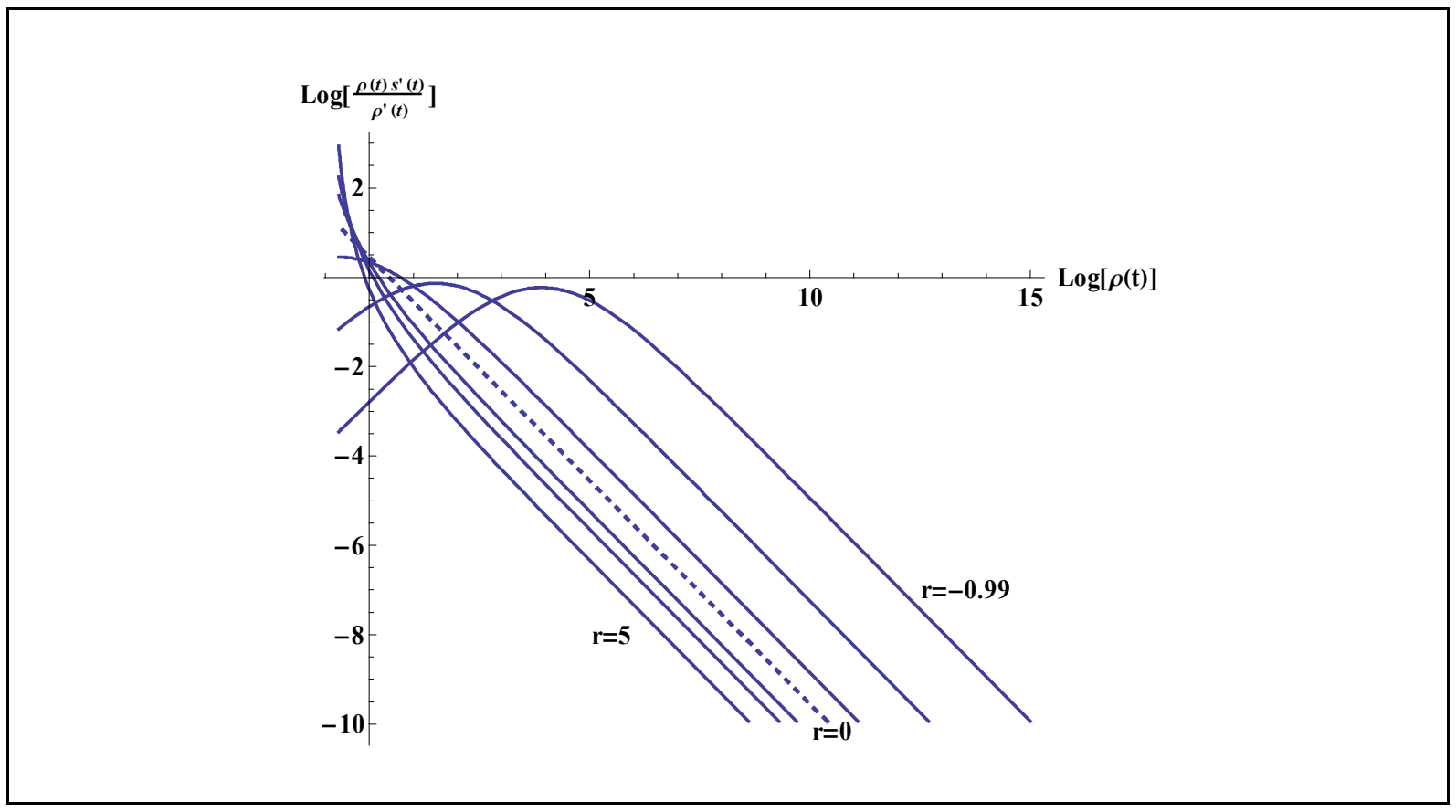

Figure 4. The dashed line is obtained when r=0 and the LCG straightens as r decreases from r=100 to r=0 and bends as r decreases further to {-0.5, -0.9,-0.99}.

Figure (5) shows the corresponding gradients of LCG for GCS segments shown in Figure (3).

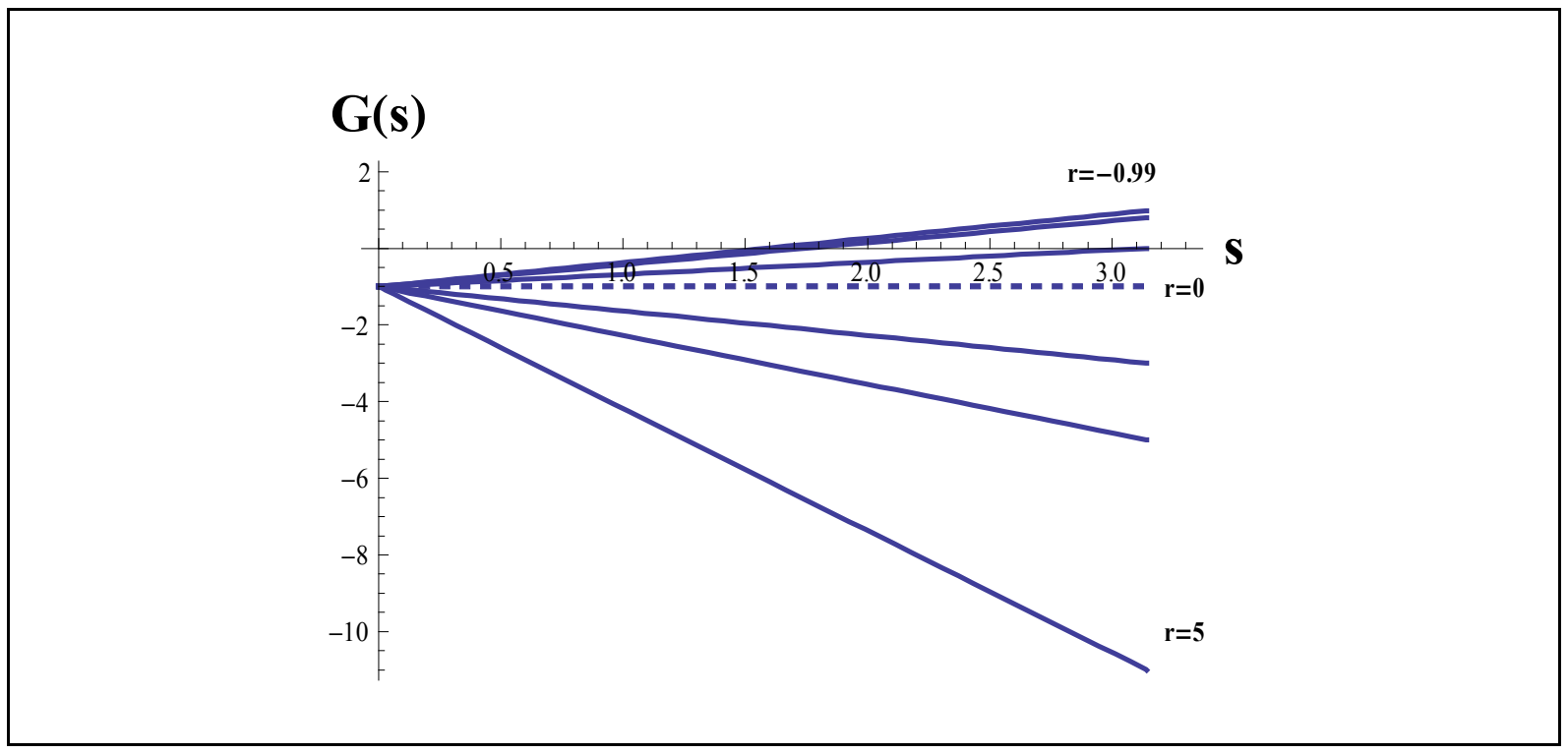

Figure 5. The horizontal dashed line is obtained when r=0 and the slope of LCG's gradient becomes negative as r increases (r=1, 2, 5) and positive as r decreases to (r=-0.5, -0.9,-0.99).

## 5. Conclusion

The only criteria Cripps [12] claimed that GCS is a high quality curve is it has a monotonic curvature profile similar to clothoid and Logarithmic spiral. In this paper, the analysis of the LCG for GCS indicated that it has linear gradient which is a general case of Log Aesthetic Curves. Hence, this paper concurs that GCS is indeed a high quality curve which can be used for practical design intent. Furthermore, these findings have led to the expansion of aesthetic curves where Gobithaasan & Miura developed first developed Generalized Log Aesthetic Curves denoted as GLAC [13]. It has been further extended as a spatial curve called Generalized Log Aesthetic Space Curve (GLASC) [14]. We strongly believe that these curves can be used as a guideline in industrial product development environment, robot trajectories planning, railway/highway design and etc.

## 6. Acknowledgements

The authors extend their gratitude to Ministry of Higher Education Malaysia for providing research grant (FRGS grant No: 59187). The authors also acknowledge anonymous referees for giving constructive comments which have helped to improve this paper.